\documentclass[aps,prb,twocolumn,showpacs]{revtex4-1}

\usepackage{graphicx}
\usepackage{latexsym}
\usepackage{amsmath}
\usepackage{amssymb}
\usepackage{amsfonts}
\usepackage{color}

\begin{document}

% Use the \preprint command to place your local institutional report
% number in the upper righthand corner of the title page in preprint mode.
% Multiple \preprint commands are allowed.
% Use the 'preprintnumbers' class option to override journal defaults
% to display numbers if necessary
%\preprint{}

%Title of paper
\title{Aharonov-Bohm oscillations in the vortex dynamics in superconducting hollow cylinders}

\author{V. N. Gladilin}
 \affiliation{TQC, Universiteit
Antwerpen, Universiteitsplein 1, B-2610 Antwerpen, Belgium}
\affiliation{INPAC, Katholieke Universiteit Leuven, Celestijnenlaan
200D, B--3001 Leuven, Belgium}
\author{J. Tempere} \affiliation{TQC,
Universiteit Antwerpen, Universiteitsplein 1, B-2610 Antwerpen,
Belgium}
\author{J. T. Devreese}
\affiliation{TQC, Universiteit Antwerpen, Universiteitsplein 1,
B-2610 Antwerpen, Belgium}
\author{V. V. Moshchalkov}
\affiliation{INPAC, Katholieke Universiteit Leuven, Celestijnenlaan
200D, B--3001 Leuven, Belgium}

\date{\today}

\begin{abstract}
Using time-dependent Ginzburg-Landau theory we demonstrate that the
Aharonov-Bohm (AB) effect, resulting from a Berry phase shift of the
(macroscopic) wavefunction, is revealed through the dynamics of
topological phase defects present in that same wavefunction. We
study vortices and antivortices on the surface of a hollow
superconducting cylinder, moving on circular orbits as they are
subjected to the force from the current flowing parallel to the
cylinder axis. Due to the AB effect the orbit deflections, caused by
a magnetic field component along the cylinder axis, become periodic
as a function of field, leading to strong and robust resistance
oscillations.
\end{abstract}

% insert suggested PACS numbers in braces on next line
% insert suggested keywords - APS authors don't need to do this
%\keywords{}

%\maketitle must follow title, authors, abstract, \pacs, and \keywords
\maketitle

% body of paper here - Use proper section commands
% References should be done using the \cite, \ref, and \label commands

\section{Introduction}
Recent advances in nanotechnology have triggered interest in
experimental and theoretical studies of mesoscopic and nanoscale
superconductivity on curvilinear
surfaces~\cite{liu01,wang05,du04a,du04b,Gladilin08,Tempere2009,lu10,Sabatino2011,Fomin2012}.
In superconducting spherical nanoshells the surface curvature leads
to a Magnus-Lorentz force, which pushes the vortices and
antivortices towards the opposite poles of the shell. This can be
considered as an effective pinning of vortices and antivortices at
the poles, which strongly affects both the equilibrium distributions
of vortices and their
dynamics~\cite{du04a,du04a,Gladilin08,Tempere2009}. The effects of
surface curvature on vortex dynamics have been recently analyzed
also for curved stripes and hollow
cylinders~\cite{Sabatino2011,Fomin2012}. An important aspect of
hollow cylinders, which have made them
 a popular subject of study, is their
doubly connected topology. The Aharonov-Bohm (AB)
oscillations~\cite{AB59}, originating from a shift of the geometric
(Berry) phase~\cite{berry} of the charge-carrier wave function by an
enclosed magnetic flux, were first experimentally observed in
superconducting hollow cylinders~\cite{deaver61} and
rings~\cite{LP62}. In the present paper, we analyze the vortex
dynamics in superconducting hollow cylinders, subjected to a
magnetic field tilted with respect to the cylinder axis. In
particular, we investigate the effect of the oscillating persistent
current, related to the AB effect, on the vortex motion and the
corresponding resistive state. The obtained results suggest that, in
parallel to the critical temperature oscillations~\cite{LP62,Wei},
also oscillations of the resistance, caused by vortex motion, can be
used as a tool to probe the AB persistent currents in
superconducting hollow cylinders.

\section{Model}

\begin{figure}
\centering
\includegraphics*[width=85 mm]{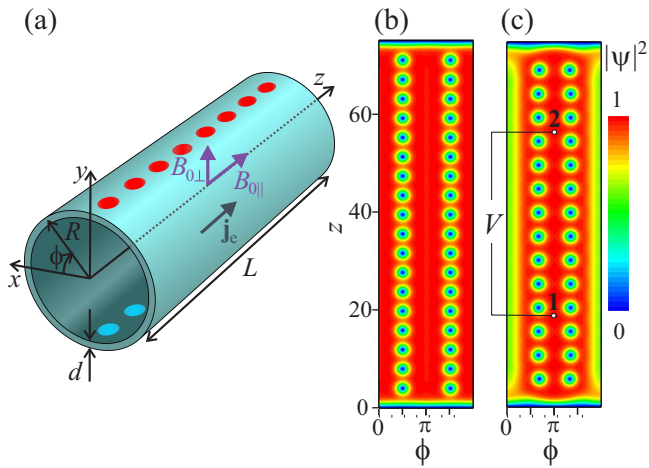}
\caption{(Color online) (a) Structure under consideration. The red
spots at $\phi=\pi/2$ (light blue spots at $\phi=3\pi/2$)
schematically show equilibrium positions of vortices (antivortices),
induced by the field $B_{0\perp}$, in the case of $j_e=0$ and
$B_{0\parallel}=0$.(b) Equilibrium distribution of the square
modulus of the order parameter in the hollow cylinder with $L = 75$,
$R = 3$, and $d \to 0$ at $B_{0\perp} = 0.35$, $B_{0\parallel} = 0$,
and $j_e = 0$ . (c) Same as in panel (b) but for $j_e = 0.22$. The
voltage drop $V$ is calculated between the points labeled as ``{\bf
1}''and ``{\bf 2}'' and located at $z=L/4$ and $z=3L/4$,
respectively. \label{Figure1}}
\end{figure}

The sketch of the structure under consideration is shown in
Fig.~\ref{Figure1}(a). A thin superconducting hollow cylinder with
thickness $d$, length $L$ and radius $R$ is subjected to an external
homogeneous magnetic field ${\bf B}_0$, which in general has nonzero
components both along the cylinder axis ($B_{0\parallel}$) and in
the perpendicular direction ($B_{0\perp}$). In the chosen
cylindrical co-ordinate frame, the $z$-axis coincides with the
cylinder axis, while the direction of $B_{0\perp}$ corresponds to
the angular coordinate $\phi = \pi/2$. At $z=0$ and $z=L$, where
normal-metal/superconducor boundary conditions are applied, an
external current with density $j_e$ is along the $z$-axis.

The vortex dynamics is described within the time-dependent
Ginzburg-Landau (TDGL) approach. For numerical simulations we use
the implementation of this approach described in detail in
Refs.~\onlinecite{Gladilin08,Silhanek11} and applicable when the
thickness of the superconductor is smaller than the Ginzburg-Landau
coherence length $\xi$, with obvious adaptations to the case of a
cylindrical layer. Like in Ref.~\onlinecite{Silhanek11}, the
relevant quantities are made dimensionless by expressing lengths in
units of $\sqrt{2}\xi$, time in units of $\pi\hbar/[4k_B(T_c-T)]
\approx 11.6\tau_{GL}$, magnetic field in units of $\Phi_0/(4\pi
\xi^2)=H_{c2}/2$, current density in units of $\Phi_0/(2\sqrt{2}\pi
\mu_0 \lambda^2 \xi)=3\sqrt{3} /(2\sqrt{2})j_c$, and scalar
potential in units of $2k_B(T_c-T)/(\pi e)$. Here,
$\Phi_0=\pi\hbar/e$ is the magnetic flux quantum, $\mu_0$ is the
vacuum permeability, $\lambda$ is the penetration depth, $\tau_{GL}$
is the  Ginzburg-Landau time, $H_{c2}$ is the second critical field,
and $j_c$ is the critical (depairing) current density of a thin wire
or film~\cite{Tinkham}. The results, described below, are obtained
for a fixed value of the Ginzburg-Landau parameter ($\kappa =
0.77$). Note that this particular choice of $\kappa$ is not really
restrictive: with the used units, the parameter $\kappa$ explicitly
enters the equations only through the expression ${\bf A}_1({\bf
r})=(2\pi\kappa^2)^{-1}\int d^3r^\prime{\bf j}({\bf r}^\prime)|{\bf
r}-{\bf r}^\prime|^{-1}$, which describes the vector potential
induced by the currents ${\bf j}$ flowing in the superconductor (see
Ref.~\onlinecite{Silhanek11} for more details). For thin
superconducting cylindrical shells under consideration ($d<1$, $d\ll
R, L$), the vector potential ${\bf A}_1$ is proportional to the
ratio $d/\kappa^2$, so that the results obtained for $\kappa = 0.77$
and a given value $d$ can be applied also to cylinders with larger
or smaller values of $\kappa$, provided that the quantity
$d/\kappa^2$ remains the same.

\section{Results and discussion}

In the absence of an applied current, the vortices and
antivortices~\cite{note}, induced in a cylindrical shell by a
magnetic field $B_{0\perp} = 0.35$, are ``geometrically pinned'' by
a Magnus-Lorentz force~\cite{Tempere2009} to $\phi = \pi /2$ and
$\phi = 3\pi /2$, respectively [see Fig.~\ref{Figure1}(b)]. When
applying a relatively low current density $j_e$ along the cylinder
in the $z$-direction, vortices and antivortices experience a force
towards $\phi=\pi$ where they can annihilate. Some of the vortices
and antivortices do annihilate with each other, but most of them
remain ``pinned'', although their equilibrium positions are somewhat
shifted towards $\phi=\pi$ due to the Lorentz force proportional to
the applied current [Fig.~\ref{Figure1}(c)]. At a certain current
density, the system enters the dissipative regime: the
vortex-antivortex pairs continuously recombine at $\phi=\pi$ while
new pairs are nucleated at $\phi=0$, where the applied current,
summed up with the Meisser current, significantly suppresses the
order parameter [see Fig.~1(c)].

\begin{figure}
\centering
\includegraphics*[width=85 mm]{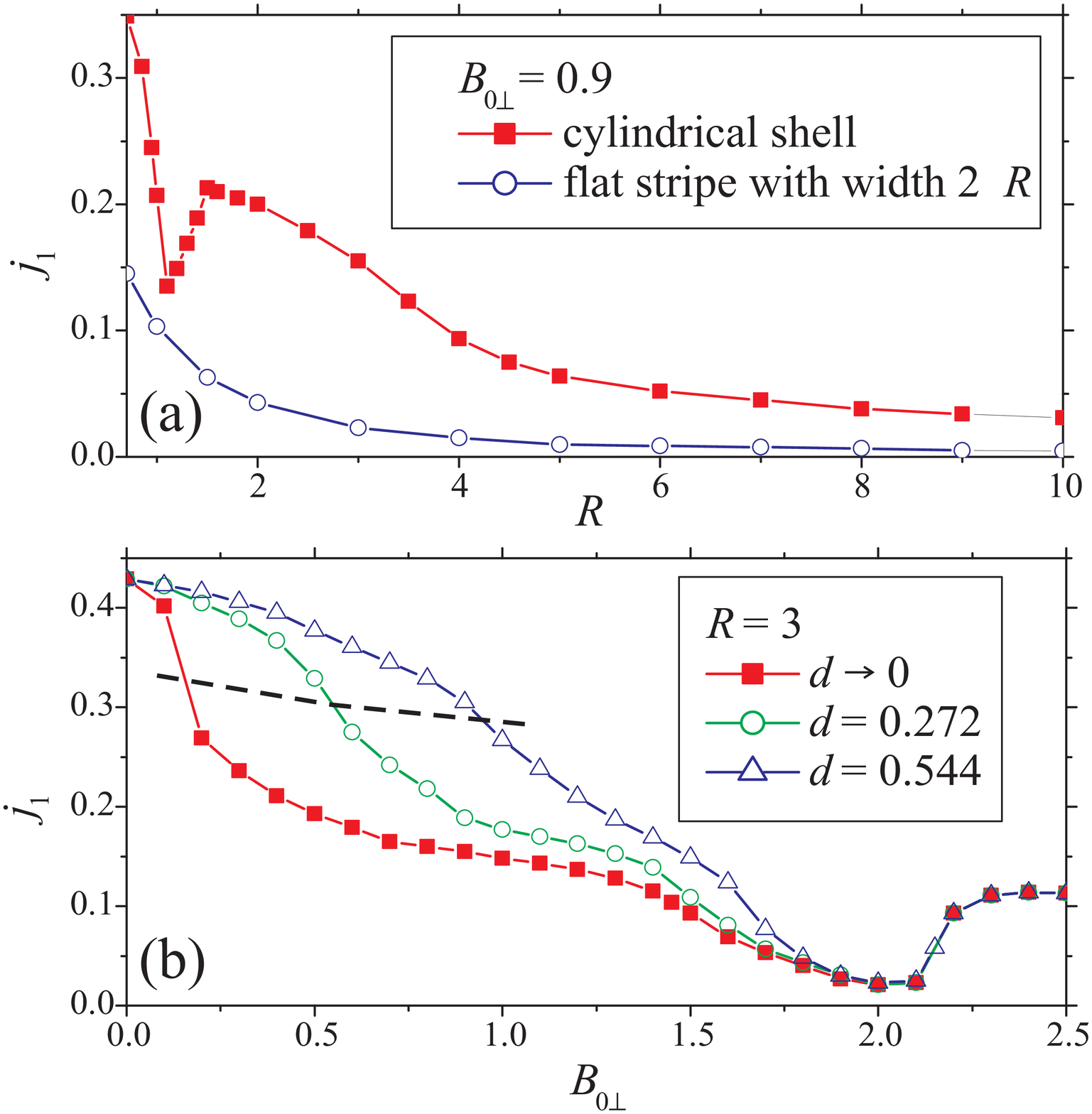}
\caption{(Color online) (a) Critical current density $j_1$ as a
function of the cylinder radius $R$ at $L = 50$, $d \to  0$,
$B_{0\perp} = 0.9$, and $B_{0\parallel} = 0$. For comparison, the
critical current density $j_1$ is shown also for a flat stripe with
width $2\pi R$, length $L = 50$ and thickness $d \to  0$ in a
perpendicular magnetic field $B_{0\perp} = 0.9$. (b) Critical
current density $j_1$ as a function of $B_{0\perp}$ at $R=3$, $L =
50$, $B_{0\parallel} = 0$, and different $d$. The dashed line
roughly shows the boundary between the ranges of parameters where
the critical current density $j_1$ corresponds to the onset of
vortex-antivortex nucleation and recombination (below the line) or
to the transition from the Meissner state to the normal state (above
the line). \label{Figure2}}
\end{figure}

In Fig.~\ref{Figure2}(a) the critical current density $j_1$,
corresponding to the onset of the resistive state in the middle part
of the cylinder (between $z=L/3$ and $z=3L/4$), is plotted as a
function of the radius $R$ of the cylindrical shell at a fixed
applied magnetic field $B_{0\perp} = 0.9$, $B_{0\parallel} = 0$. As
seen from Fig.~\ref{Figure2}(a), the critical current density
$j_1(R)$ for cylindrical shells is appreciably higher than that for
the corresponding flat stripes with the same cross-section and
manifests a nonmonotonous behavior. As the radius is decreased (down
to $R \approx 1.5$), the critical current density increases. This is
due to an enhancement of the geometric pinning of vortices and
antivortices as $R$ gets smaller. However, for $1.1 <R < 1.5$ the
behavior of $j_1(R)$ becomes qualitatively different: it decreases
with reducing $R$. This decrease is related to the fact that at
those radii the distance between vortices, located at $\phi\approx
\pi/2$, and antivortices, located at $\phi\approx 3\pi/2$, becomes
comparable to the vortex size, so that their annihilation is
facilitated by reducing $R$. At $R < 1.1$ even rather weak external
current densities $j_e$ are sufficient for complete annihilation of
vortex-antivortex pairs. In this case, the onset of a resistive
state is determined by the nucleation of new pairs, which requires
larger applied current densities $j_e$ at smaller $R$. As a result,
in this range of $R$ the critical current density rapidly increases
with decreasing $R$.

The calculated dependence of $j_1$ on the applied field $B_{0\perp}$
[Fig.~\ref{Figure2}(b)] is qualitatively similar to that found in
Ref.~\onlinecite{Sabatino2011} in the limit $d \to 0$ and for
applied fields below $H_{c2}$ ($B_{0\perp}<2$). Increasing the shell
thickness $d$ leads to an increase in the critical current density
$j_1$ due to partial screening of the applied field $B_{0\perp}$ by
supercurrents, so that a higher current must be applied in order to
produce a Lorentz force sufficient for vortex/antivortex depinning.
The shape of the curves $j_1(B_{0\perp})$ reveals several regimes.
(i) At relatively low magnetic fields, no vortices appear in the
cylinder, so that the values of $j_1$ actually correspond to the
transition from the Meissner state to the normal state. (ii) With
increasing applied magnetic field, the critical current density for
nucleation of vortex-antivortex pairs, becomes smaller than the
depairing current density so that $j_1$ corresponds to the onset of
vortex-antivortex nucleation, propagation, and annihilation. (iii)
At even higher fields, vortices and antivortices are present in the
cylinder already at $j_e < j_1$. The critical current density
decreases monotonously with increasing $B_{0\perp}$ [both due to an
increase of the depinning Lorentz force, proportional to
$B_{0\perp}$, and an enhanced mutual repulsion of the increasing
number of vortices (antivortices)]. (iv) When approaching the second
critical field ($B_{0\perp}=2$), the number and density of vortices
(antivortices) become so large that the distances between the vortex
cores are smaller than the vortex size. The resulting enhancement of
mutual repulsion between vortices (antivortices) leads to a rather
fast decrease of $j_1$ with $B_{0\perp}$ in this field range. At
$B_{0\perp}=2$ the curves $j_1(B_{0\perp})$ exhibit a pronounced
minimum. (v) At $B_{0\perp}>2$, the critical current density first
sharply increases and then slowly decreases. In this field range,
normal regions are formed in the cylinder around  $\phi = \pi/2$ and
$\phi = 3\pi/2$, and the vortex dynamics involves the entrance of
vortices and antivortices from the normal regions to the
superconducting regions, which requires relatively high applied
current densities. Obviously, only the range of parameters
corresponding to regimes (ii) to (iv) can be relevant for revealing
the effect of persistent currents, induced by the field
$B_{0\parallel}$, on vortex dynamics. As illustrated by
Fig.~\ref{Figure2}, this range is sufficiently wide.

\begin{figure}
\centering
\includegraphics*[width=85 mm]{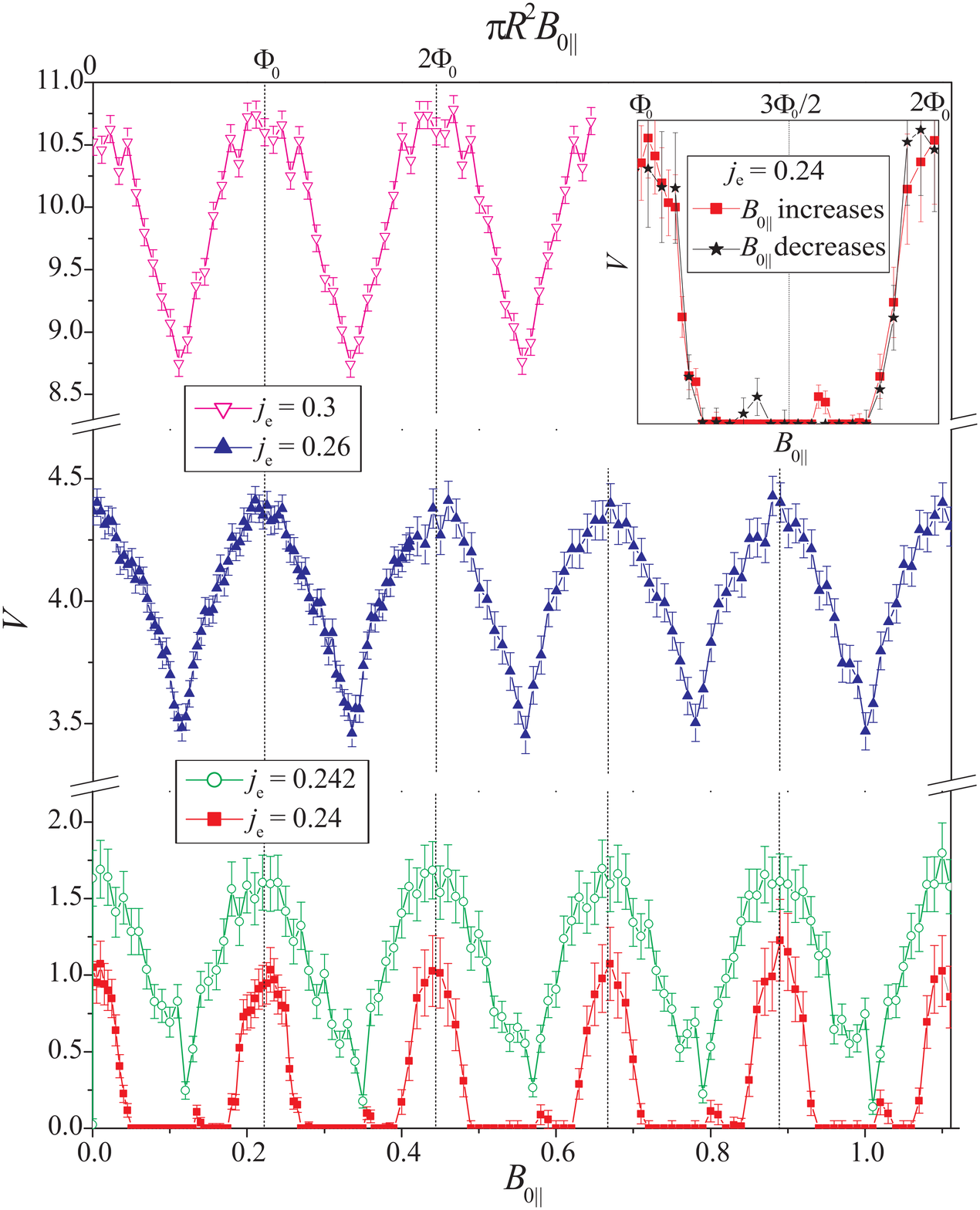}
\caption{(Color online) Time-averaged voltage drop $V$ between
points 1 and 2 [see Fig.~1(c)] in a hollow cylinder with $L = 75$,
$R = 3$, and $d \to 0$ as a function of an increasing magnetic field
$B_{0\parallel}$ for $B_{0\perp} = 0.35$ and different applied
current densities $j_e$. Inset: Voltage drop $V$ as a function of
the applied parallel magnetic field in the cases of increasing
(squares) and decreasing (stars) field $B_{0\parallel}$ for $L =
75$, $R = 3$, $d \to 0$, and $B_{0\perp} = 0.35$.  \label{Figure3}}
\end{figure}

In Fig.~\ref{Figure3} we plot the calculated time-averaged voltage
drop $V$ between points 1 and 2 [see Fig.~\ref{Figure1}(a)] as a
function of an increasing parallel magnetic field $B_{0\parallel}$
for different values of the externally applied current density
$j_e$. The accuracy of the shown $V$-values is determined by a
limited averaging time in the performed calculations. When switching
$B_{0\parallel}$ to a new value, a time interval $t_{\bf r}$,
ranging from 100 to 500, is reserved in the computational program
for transient processes. Within this time interval, no calculation
of $V$ is performed. Then the voltage drop $V$ is averaged over a
time interval $t_{\bf aver}$, ranging from 500 to 5000 in the
present calculations. However, our analysis shows that in the
multi-vortex system under consideration the full period of
vortex-antivortex generation/recombination processes can be rather
long, so that in general this period is not much smaller than the
used time intervals $t_{\bf aver}$. The error bars, shown in
Fig.~\ref{Figure3}, correspond to the estimates, obtained using
different values of $t_{\bf r}$ and $t_{\bf aver}$.

\begin{figure}
\centering
\includegraphics*[width=80 mm]{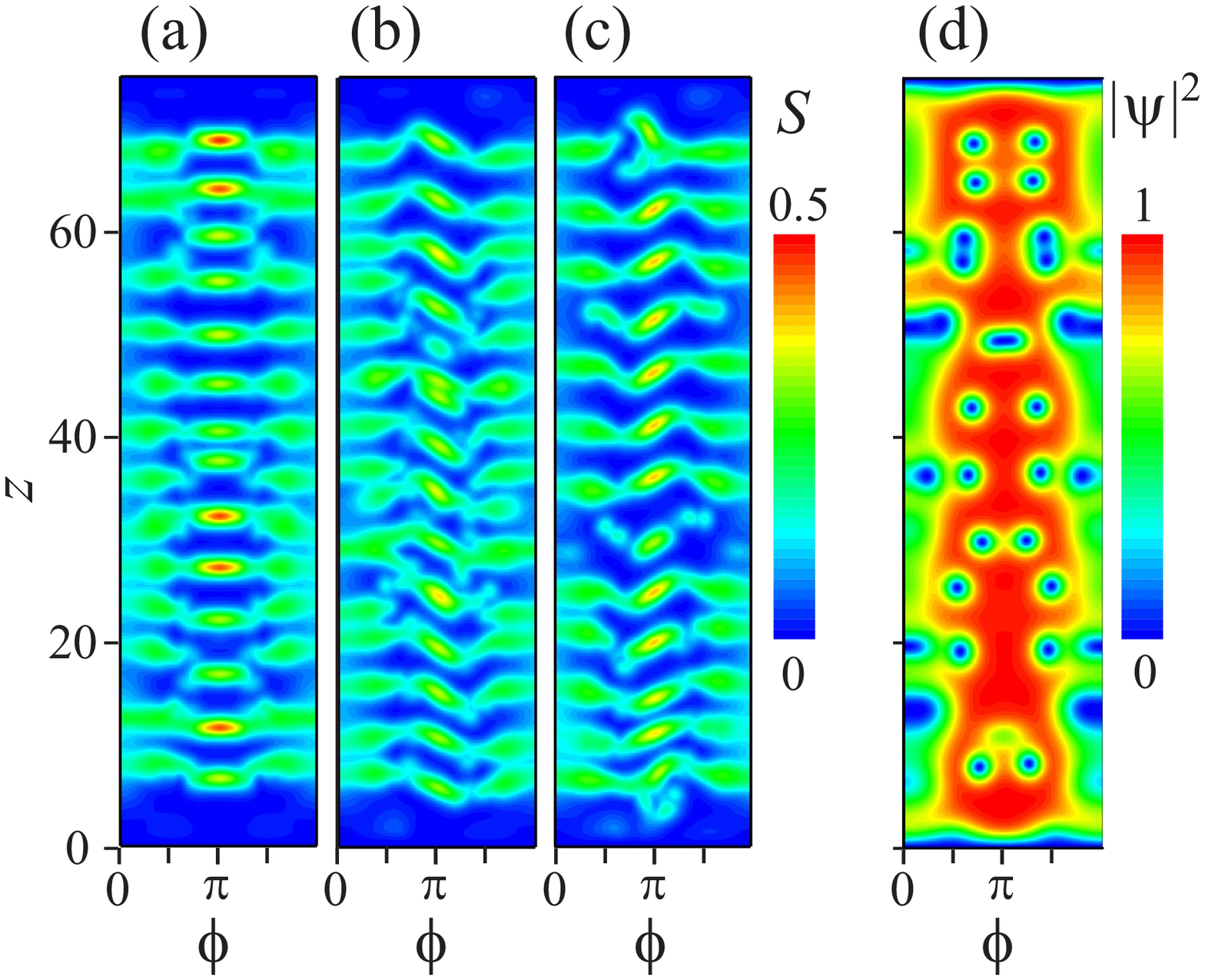}
\caption{(Color online) Panels (a) to (c): Distributions of the
streaming parameter $S$  for $L = 75$, $R = 3$, $d \to 0$,
$B_{0\perp} = 0.35$, $j_e = 0.26$, $t_2- t_1 = 15$ and different
values of the applied parallel magnetic field: $B_{0\parallel}=0$
(a), $B_{0\parallel}=0.1$ (b), and $B_{0\parallel}=0.125$ (c). Panel
(d): Snapshot of the square modulus of the order parameter in the
cylindrical shell with $L = 75$, $R = 3$, and $d \to 0$ at
$B_{0\perp} = 0.35$, $j_e = 0.3$, and $B_{0\parallel}=0.2$.
\label{Figure4}}
\end{figure}

As seen from Fig.~\ref{Figure3}, there are well pronounced
oscillations of the calculated voltage $V$ versus $B_{0\parallel}$.
The oscillation period is typical for the AB oscillations and equals
$\Phi_0/(\pi R^2)$. The voltage $V$ takes maximum values for
vanishing persistent currents induced by the field $B_{0\parallel}$,
i.e. for $\pi R^2 B_{0\parallel}\approx (2n+1)\Phi_0/2$
($n\in\mathbb{Z}$), while the sharp minima of $V$ approximately
correspond to the maximum magnitude of the persistent currents, i.e.
to $\pi R^2 B_{0\parallel}= (n+1/2)\Phi_0$ ($n\in\mathbb{Z}$). In
other words, the shape of these oscillations is ``inverted'' as
compared to the Little-Parks oscillations~\cite{LP62}. The
oscillating behavior of $V(B_{0\parallel})$, shown in
Fig.~\ref{Figure3}, can be explained in terms of the distortion of
vortex/antivortex trajectories due to the Lorentz forces caused by
the persistent currents and the magnetic field $B_{0\perp}$. In
Figs.~\ref{Figure4}(a) to \ref{Figure4}(c), we plot the
distributions of the streaming parameter
$S=\left[({t_2}-{t_1})^{-1}\int_{t_1}^{t_2} (\partial
|\psi|^2/\partial t)^2dt\right]^{1/2}$, introduced in [2] to
visualize the vortex/antivortex motion. As implied by
Fig.~\ref{Figure4}(a), in the case of $B_{0\parallel}=0$, and hence
in the absence of the corresponding persistent currents, the
vortex/antivortex trajectories in the middle part of the cylinder
follow a circular cross-section of the cylinder, with vortices
moving from $\phi=0$ to $\pi$ and antivortices moving from $\phi=2
\pi$ to $\pi$ all at the same value of $z$. Vortex-vortex
interactions cause only slight deviations from ideal circles.
However, in the cases when the field $B_{0\parallel}$ induces strong
diamagnetic [$B_{0\parallel} = 0.1$; see Fig.~\ref{Figure4}(b)] or
paramagnetic [$B_{0\parallel} = 0.125$; see Fig.~\ref{Figure4}(c)]
currents in the cylindrical shell, these currents cause a rather
pronounced deformation of vortex/antivortex trajectories, so that
the path of a vortex (antivortex) from $\phi=0$ ($2\pi$) to
$\phi=\pi$ partly follows the ellipse traced out by a slanted
cross-section of the cylinder. This represents an appreciably longer
path than for $B_{0\parallel}=0$.

The behavior of $V(B_{0\parallel})$ at the applied current density
$j_e = 0.24$, which only slightly exceeds the critical current
density $j_1 = 0.2392$ at $B_{0\parallel}=0$, implies that -- in
addition to oscillations of $V(B_{0\parallel})$ at a fixed
$j_e$-value -- also the critical current density $j_1$ is an
oscillating function of the applied parallel magnetic field
$B_{0\parallel}=0$. However, our calculations show that the
corresponding oscillation amplitudes are very small. For the
parameters under consideration the critical current at $\pi R^2
B_{0\parallel}= (n+1/2)\Phi_0$ is $j_1 = 0.2416$, i.e. the
oscillation amplitude for $j_1$ does not exceed 0.6\%. Thus we
conclude that the voltage $V(B_{0\parallel})$ is a better indicator
than the critical current to reveal the AB oscillations in vortex
trajectories.

Besides pronounced oscillations of the voltage $V(B_{0\parallel})$
with period  $\Phi_0/(\pi R^2)$, one can see in Fig.~\ref{Figure3}
several smaller features in the behavior of $V(B_{0\parallel})$.
Many of those features are irregular and reflect the limited
computation accuracy. However, some others are rather regular and
periodic in $B_{0\parallel}$. In particular, for $j_1 = 0.24$, small
peaks of $V(B_{0\parallel})$ emerge at magnetic fields slightly
above $B_{0\parallel}= (n+1/2)\Phi_0/(\pi R^2 )$. Those features can
be attributed to the fact that the (time-averaged) number of
vortex-antivortex pairs in the cylinder depends not only on the
applied current density [see Figs.~\ref{Figure1}(b) and
\ref{Figure1}(c)] but also on the parallel field $B_{0\parallel}$.
At the same time, the time-averaged voltage drop $V$ as well as the
critical current $j_1$ are, of course, sensitive to the number of
vortices in the cylinder. This sensitivity leads, in particular, to
the appearance of the peaks of $V(B_{0\parallel})$ for $j_e = 0.24$
and $B_{0\parallel}$ slightly above $(n+1/2)\Phi_0/(\pi R^2 )$.

From Fig.~\ref{Figure3}, one can also observe that the curves
$V(B_{0\parallel})$ are not fully symmetric with respect to the
points $B_{0\parallel}= n\Phi_0/(\pi R^2 )$ or $B_{0\parallel}=
(n+1/2)\Phi_0/(\pi R^2 )$. Thus, in Fig.~\ref{Figure3}, the
aforementioned weak peaks of $V(B_{0\parallel})$ at $j_1 = 0.24$ are
present at magnetic fields just above $B_{0\parallel}=
(n+1/2)\Phi_0/(\pi R^2 )$, but no similar peaks appear below
$B_{0\parallel}= (n+1/2)\Phi_0/(\pi R^2 )$. In our numerical
simulations, we ramp up the magnetic field $B_{0\parallel}$ and
calculate how the order parameter adapts to the increased field.
Geometric pinning of vortices in the cylinder leads to hysteresis
effects and results in asymmetries in the curves. A similar pinning
in superconducting spherical nanoshells has been shown to cause a
pronounced hysteresis in the dependence of the number of
vortex-antivortex pairs on the applied magnetic
field~\cite{Tempere2009}. Also in the cylindrical shells under
consideration the number of vortex-antivortex pairs and hence the
curves $V(B_{0\parallel})$ demonstrate hysteretic behavior. The
presence of such a hysteretic behavior is illustrated in the inset
of Fig.~\ref{Figure3}, where we compare the dependences
$V(B_{0\parallel})$, calculated for increasing and decreasing fields
$B_{0\parallel}$. As seen from this inset, in the case of a
decreasing field $B_{0\parallel}$, small peaks of
$V(B_{0\parallel})$ appear to the left from $B_{0\parallel}=
(3/2)\Phi_0/(\pi R^2 )$. Within the error bars, the whole pattern of
$V(B_{0\parallel})$, where the field $B_{0\parallel}$ goes either up
or down, looks symmetric with respect to $B_{0\parallel}=
(3/2)\Phi_0/(\pi R^2 )$.

As further seen from Fig.~\ref{Figure3}, the shape of the curve
$V(B_{0\parallel})$ becomes more regular and symmetric when
increasing the applied current density to the value $j_e=0.26$,
which is considerably higher than the critical current density
$j_1$. However, relatively pronounced additional features, caused by
variations of the (time-averaged) number of vortex-antivortex pairs
in the cylinder, reappear in the curve $V(B_{0\parallel})$ at
$j_e=0.3$ (Fig.~\ref{Figure3}), when -- as illustrated by
Fig.~\ref{Figure4}(d) -- some precursors of phase-slip-line
formation can be already seen in the order parameter pattern.
Remarkably, within the whole range of current densities considered
here, the oscillation amplitude of $V(B_{0\parallel})$ remains
appreciably large.

\begin{figure}
\centering
\includegraphics*[width=80 mm]{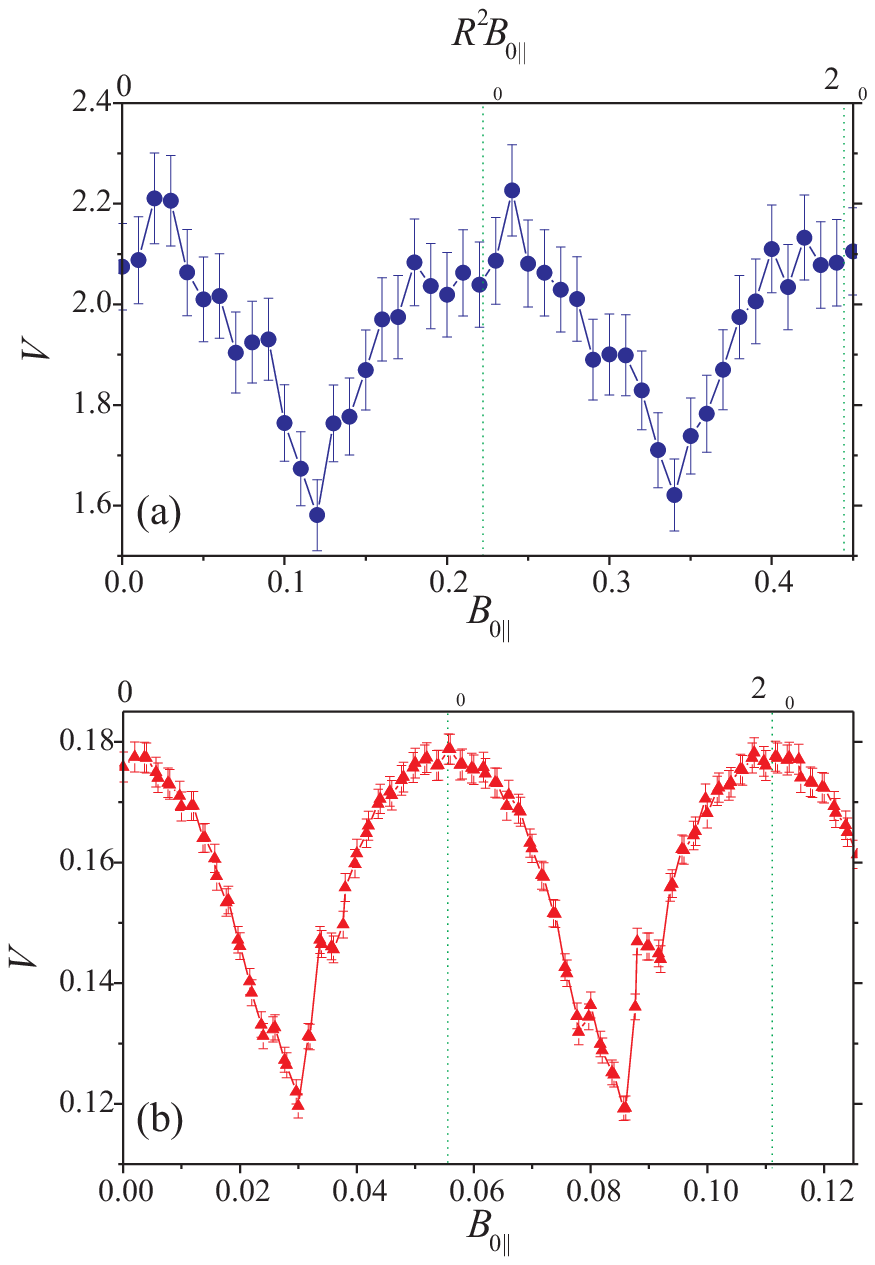}
\caption{(Color online) Time-averaged voltage drop $V$ as a function
of an increasing magnetic field $B_{0\parallel}$ in the cases of (a)
$L = 30$, $R = 3$, $d =0.272$, $B_{0\perp} = 0.6$, $j_e=0.28$ and
(b) $L = 30$, $R = 6$, $d \to 0$, $B_{0\perp} = 0.2$, $j_e=0.2$.
\label{Figure5}}
\end{figure}

From Fig.~\ref{Figure3}, the magnitude of the resistivity
oscillations is $\geq 0.05$ in the used units. This is about 0.5\%
of the normal-state resistivity, which equals 12 in our
units~\cite{Silhanek11}, so that the predicted oscillations should
be observable through 4-probe measurements similar to those reported
in Ref.~\onlinecite{wang05}. Our calculations show that the
self-inductance of the cylinders under consideration has a
relatively weak effect on the resistivity oscillations; as follows
from Fig.~\ref{Figure5}(a), for $L=30$, $R=3$, $d=0.272$,
$\kappa=0.77$, $B_{0\perp} = 0.6$ and $j_e=0.28$ the oscillation
magnitude is about 0.07. This magnitude tends to decrease when
increasing the radius of the cylinder and/or when decreasing its
length down to values $L < 2\pi R$ [cp. Fig.~\ref{Figure5}(b) to
Fig.~\ref{Figure3}]. Nevertheless, for $L=30$, $R=6$, $d\to 0$,
$B_{0\perp} = 0.2$ and $j_e=0.2$ our calculations predict an
oscillation magnitude as large as 0.01[see Fig.~\ref{Figure5}(b)].
Of course, in very short cylinders with $R\sim 1$, which cannot
accommodate vortices, the predicted voltage oscillations become
impossible.

It seems worth emphasizing that the material and geometric
parameters, required to observe the predicted resistivity
oscillations, are achievable experimentally. For example, with
$\xi\approx 150$~nm (close to the values of $\xi(0)$ for Al hollow
cylinders in Ref.~\onlinecite{wang05}), the dimensionless parameters
$R = 3$ and $d = 0.272$ would correspond to a hollow cylinder with
radius about 640 nm (a few times larger than the cylinder radii in
Ref.~\onlinecite{wang05}) and wall thickness about 60 nm (twice that
in Ref.~\onlinecite{wang05}).

\section{Conclusions}

To conclude, we have shown that in hollow superconductor cylinders,
subjected to a tilted magnetic field, the resistance, caused by
vortex motion, should manifest measurable oscillations as a function
of the magnetic field component parallel to the cylinder axis. This
effect can provide a robust tool to probe experimentally the
oscillating persistent currents, related to the Aharonov-Bohm
effect, in a wide range of parameters, in particular, much below the
superconducting critical temperature.

\begin{acknowledgments}
This work was supported by Methusalem funding by the Flemish
government, the Flemish Science Foundation (FWO-Vl), in particular
FWO projects G.0356.05, G.0115.06, G.0370.09N, and G.0115.12N, the
Scientific Research Community project WO.033.09N, the Belgian
Science Policy, and the ESF NES network.
\end{acknowledgments}

% Create the reference section using BibTeX:

\end{document}